\documentclass[graybox]{svmult}

\usepackage{mathptmx}       
\usepackage{helvet}         
\usepackage{courier}        
\usepackage{type1cm}        
\usepackage{makeidx}         
\usepackage{graphicx}        
\usepackage{multicol}        
\usepackage[bottom]{footmisc}

\usepackage{amsfonts,amssymb,amsmath,bm,bbm}
\usepackage{graphicx,epsfig,color}
\usepackage{url}
\usepackage{comment}
\usepackage{ulem}

\makeindex

\begin{document}

\title{Ensemble Control of Cycling Energy Loads: Markov Decision Approach}
\titlerunning{Ensemble Control of Loads}

\author{Michael Chertkov, Vladimir Y. Chernyak and {Deepjyoti Deka}}
\authorrunning{Michael Chertkov, Vladimir Chernyak and Deepjyoti Deka}
\institute{
Michael Chertkov \at Theoretical Division, T-4 \& CNLS,
Los Alamos National Laboratory
Los Alamos, NM 87545, USA and
Energy System Center, Skoltech, Moscow, 143026, Russia, \email{chertkov@lanl.gov} \and
Vladimir Y. Chernyak \at Department of Chemistry, Wayne State University, 5101 Cass Ave, Detroit, MI 48202, USA, \email{chernyak@chem.wayne.edu} \and
{Deepjyoti Deka \at Theoretical Division, T-4 \& CNLS,
Los Alamos National Laboratory
Los Alamos, NM 87545, USA, \email{deepjyoti@lanl.gov}
}}

\maketitle

\abstract{A Markov decision process (MDP) framework is adopted to represent ensemble control of devices with cyclic energy consumption patterns, e.g., thermostatically controlled loads. Specifically we utilize and develop the class of MDP models previously coined linearly solvable MDPs, that describe optimal dynamics of the probability distribution of an ensemble of many cycling devices. Two principally different settings are discussed. First, we consider optimal strategy of the ensemble aggregator balancing between minimization of the cost of operations and minimization of the ensemble welfare penalty, where the latter is represented as a KL-divergence between actual and normal probability distributions of the ensemble.  Then, second, we shift to the demand response setting modeling the aggregator's  task to minimize the welfare penalty under the condition that the aggregated consumption matches the targeted time-varying consumption requested by the system operator. We discuss a modification of both settings aimed at encouraging or constraining the transitions between different states. The dynamic programming feature of the resulting modified MDPs is always preserved; however, `linear solvability' is lost fully or partially, depending on the type of modification. We also conducted some (limited in scope) numerical experimentation using the formulations of the first setting. We conclude by discussing future generalizations and applications.}

\section{General Motivation/Introduction}

Power systems, as well as other energy systems, have undergone a transition from traditional device-oriented and deterministic approaches to a variety of novel approaches to account for
\begin{itemize}
\item stochasticity and uncertainty in how devices, especially new devices such as wind farms, are operated;
\item network aspects, e.g., with respect to optimization, control, and design/planning; and
\item utilization of massive amounts of newly available measurements/data for the aforementioned settings.
\end{itemize}
Such approaches, in particular Demand Response (DR), have become one important component of smart grid development, see \cite{11CH,14Siano} and references therein. Novel DR architectures break the traditional paradigm where only generators are flexible, and hence suggest that participation of flexible loads in control can benefit the grid at large without compromising load/consumer comforts significantly.

DR assumes that the loads are capable of following operational commands from the system operator (SO). Using DR in the range from tens of seconds to minutes is a potential attractive niche for frequency control that maintains the balance between production and consumption \cite{80STKOPC,10DHMT,13AKCAS,16ZHZM}. Traditional frequency control is achieved by adjusting the generators,  whereas DR (when developed) helps to achieve the balance additionally by adjusting the loads. The control task for loads in the DR setting is set by the SO as a temporal consumption request. Such requests can be formulated ahead of time (e.g., for the next 10 min) or in real time, using  frequent updates (e.g., every 10 seconds).

This type of frequency control, namely DR services, is already provided by big consumers such as aluminum smelters \cite{09Alcoa} and desalination plants \cite{16DesERCOT}. However, the potential effect would be an order of magnitude larger if small loads are available to provide DR services. Nevertheless, direct involvement of small-scale consumers is expensive because of associated communication and control costs. A viable solution to this problem is to control many small-scale consumers indirectly via an intermediate entity, also called the aggregator of an ensemble that includes many (e.g., thousands or tens of thousands) small-scale consumers \cite{79CD,81IS,84CM,85MC,88MC,04LC,05LCW,09Cal,11CH,11BF}. Therefore, this novel DR architecture assumes, action-wise, separation into the following three levels:
\begin{itemize}
\item [(a)] SO sends control request/signal to an aggregator. The SO request is stated as a temporal profile (possibly binned into {short intervals e.g. 15 seconds) that is expected from the ensemble for an upcoming duration of length e.g. 10 minutes.}

\item [(b)] Aggregator (A) processes the SO signal by solving an optimization/control problem and broadcasts the same A-signal, which is the output/solution of the optimization/control problem, to all members of the ensemble (end-point consumers). The broadcast of identical signal to end-point devices/consumers, makes the communication step cheap.

\item [(c)] End-point devices receive and implement the A-signal into control action. The implementation is assumed to be straightforward, and at most require only light device-level computations.
\end{itemize}

A number of challenges are associated with this novel aggregator-based DR architecture. The challenges are of both formulation (conceptual) and solution (implementation) type. We will mention a couple of these challenges most relevant to level (b) of the control architecture, which is the focus of this manuscript.

{In posing the A-optimization, one desires a simple/minimal, yet sufficiently accurate, way of modeling individual devices. In particular, an acceptable framework needs to model the devices in terms of their achievable states and spatio-temporal resolution. To address this challenge, we will state the aggregator-level problems through the language of Markov decision process (MDP), and therefore utilize the transition probabilities between states in MDP as control variables. More accurately, we will assume that each device can find itself in all (or a subset of all) of the finite number of states. Then we describe the probabilistic state of the ensemble in terms of a vector of non-negative numbers that represent the fraction of all consumers observed in different states at any given moment of time. At each time, the probabilistic state vector thus sums to unity.} The optimization/control degree of freedom for the A-optimization is represented by the set of stochastic transition probability matrices between the states, defined at each time slot over the discretized and finite time horizon.

In addition to the frequency control formulation, we will also discuss another setting that is of interest by itself but also less challenging in terms of finding efficient computational solutions. In this algorithmically simplified case we will look for an optimal balance between the cost of the ensemble operations when the price of electricity changes in time and the deviation of the probabilistic state of the ensemble from its natural/normal behavior under uniform prices or without the price bias.

{The discrete-time, discrete-space ensamble modeling has a continuous-time, mixed-space counterpart known as thermostatically controlled loads (TCLs), or even more generally cycling loads, that are characterized by periodic (or quasi-periodic) evolution in the phase space.} See \cite{11CH} and our recent paper \cite{17CCa} for detailed discussions of TCL modeling and relevant references. In this manuscript we choose to work with discrete-time, discrete-space Markov process (MP) models because of their universality and flexibility. Indeed, an MP model can be viewed as a macro-model that follows from its micro-model counterpart - the TCL, after spatio-temporal model reduction (coarse-graining), see e.g. \cite{15PKL}. However, MPs can represent a much broader class of models than those obtained via coarse-graining of TCLs or even than a combination (average) of different TCL models. The  more general class of MDP models is especially useful in the context of machine learning (ML), where an MP is reconstructed from actual measurements/samples of the underlying ensemble.

From MP, which represents stochastic dynamics, we transition to MDP, which represents stochastic optimization with the MP/ensemble state constrained to being within a specified class of state dynamics. Many choices of MDP are reasonable from the perspective of practical engineering. In this manuscript we adopt and develop an approach pioneered in \cite{82FM} and further developed in \cite{05Kap,05Meyn,07Tod,12DjEmo,13DjEmo,15MBBCE}. This relatively unexplored formulation of the stochastic optimal control is known as ``path integral control" in the case of continuous-time, continuous-space formulations \cite{05Kap} and as ``Linearly Solvable MDP" (LS-MDP) in the case of a discrete-time, discrete-space setting \cite{07Tod,12DjEmo,13DjEmo}. Linear solvability is advantageous because it allows us to determine analytic expressions for the optimal solution in settings that go one step further than possibly through Dynamic Programming (DP) approaches.  (We remind the reader that DP involves recursive solutions of the Hamilton--Jacobi--Bellman [HJB] equations.) One extra advantage of the LS-MDP, which holds even when reduction of the HJB equation to a linear equation is not possible, is related to the fact that the penalty term associated with deviation of the optimized transition probability from its ideal shape (not perturbed by the SO signal) has a very natural form of the generalized Kublack--Leibler (KL) distance \cite{coverthomas} between the two probability measures.

\begin{figure}
\centering
\includegraphics[scale=0.37,page=2]{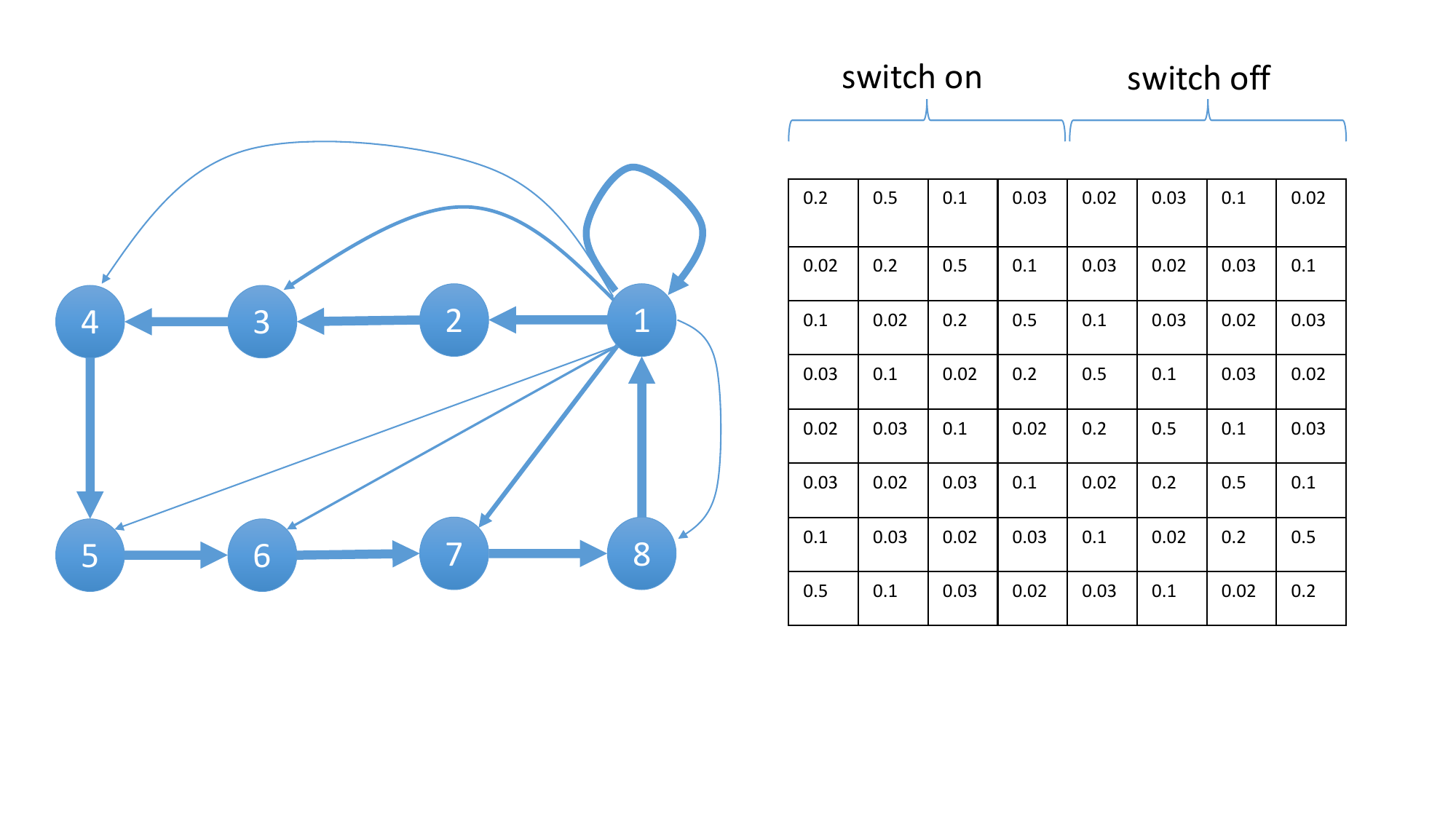}
\caption{An example MDP resulting from space-binning/discretization and time-discretization of a two-level mixed-state TCL model of the type discussed in \cite{17CCa}. This model is used in Section \ref{sec:experiments} for illustrative experiments. The directed graph of the ``natural" transitions and respective stochastic $\bar{p}$ matrix are shown in the left and right subfigures, respectively.
\label{fig:TCL_MDP}}
\end{figure}

When discussing MPs and MDPs, we will utilize the following (rather standard) notations and terminology:
\begin{itemize}
\item The dynamic state of the ensemble is described in terms of a vector, $\rho(t)=(\rho_\alpha(t)|\forall \alpha)$, where $\rho_\alpha(t)\geq 0$ is the probability to find a device (from the ensemble) in state $\alpha$ at time $t$. The vector is normalized (no probability loss) at all times considered, i.e., $\sum_\alpha \rho_\alpha(t)=1,\quad\forall t{=0 ,\cdots, T},~~\forall \alpha$.
\item $p(t)=(p_{\alpha\beta}(t)|\forall t,\forall \alpha,\beta)$ is the transition probability matrix that describes an MP. It is set to be an optimization variable within the MDP framework. The matrix element, $p_{\alpha\beta}(t)$, represents the transition probability, i.e. the probability for a device which was at state $\beta$ at time $t$ to transition to state $\alpha$ at $t+1$. The matrix is stochastic, i.e.
\begin{eqnarray}
\sum_\alpha p_{\alpha\beta}(t)=1,\quad \forall t=0 ,\cdots, T-1,~~ \forall \beta,
\label{stochastic}
\end{eqnarray}
\item The evolution of the MP/ensemble state is described by the master equation (ME):
\begin{eqnarray}
\rho_\alpha(t+1)=\sum_\beta p_{\alpha\beta}(t)\rho_\beta(t),\quad \forall t,~~ \forall \alpha.
\label{master-eq}
\end{eqnarray}
The ME should be supplemented by the initial condition for the state of the ensemble:
\begin{eqnarray}
\rho_\alpha(0)=\rho_{in;\alpha},\quad \forall \alpha
\label{rho_init}
\end{eqnarray}
Given the initial condition (\ref{rho_init}), one needs to run ME (\ref{master-eq}) forward in time to find the states of the MP ensemble at all future times. The direction of time is important because it reflects the physical causality of the setting.

\item MDP formulations discussed in the manuscript 
are stated as optimizations over the transition probability matrix $p$. All of the formulations also depend on the target $\bar{p}$, which corresponds to the optimal transition matrix if the ensemble is ``left alone" for a sufficiently long time. In the case of frequency control, discussed in Section \ref{sec:freq_control}, $\bar{p}$ corresponds to the steady state of the ensemble when the frequency tracking guidance is ignored. In the formulations of
Sections \ref{sec:profit_optimal} and \ref{sec:penalty}, $\bar{p}$ corresponds to the steady state of the ensemble operating for a sufficiently long time in the case of a flat, i.e., time- and state-independent, cost. An example MP and respective ``natural" $\bar{p}$, introduced in the result of coarse-graining of a TCL, is shown in Fig.~(\ref{fig:TCL_MDP}).
\end{itemize}

The material in the remainder of the manuscript is organized as follows. MDPs aimed at finding the profit optimal transition probability vector for an ensemble of devices are defined and analyzed in Section \ref{sec:profit_optimal}. MDPs discussed in this section measure the deviation of the ``optimal" $p$ from its ``normal" counterpart, $\bar{p}$, in terms of the standard KL divergence. {This formulation is an LS-MDP; we have placed a detailed technical discussion of the MDP's linear solvability and related features and properties in Appendix \ref{app:DP}. Next, in Section \ref{sec:penalty}, a modified MDP is discussed that differentiates between state transitions, i.e., discounts some transitions and encourages others. This formulation is richer in comparison to the differentiation-neutral formulation of Section \ref{sec:profit_optimal}. Numerical simulations for the optimization objectives discussed in Sections  \ref{sec:profit_optimal} and \ref{sec:penalty} are presented in Section \ref{sec:experiments}. In Section \ref{sec:freq_control} we describe and discuss the solution of an MDP, seeking to minimize the welfare deviation, stated in terms of the KL distance (or weighted KL distance) between the ``optimal'' and ``normal'' transition probabilities while also matching the SO objective exactly. Preliminary discussion on integrating the MDP approaches into the optimum power flow formulations, e.g., to jointly control grid voltages and to minimize the power losses, is given in
Section \ref{sec:voltage}.} Section \ref{sec:conclusions} is reserved for the conclusions and discussions of the path forward.

\section{Cost-vs-Welfare Optimal}
\label{sec:profit_optimal}

Our main MDP formulation of interest, which we term ``cost vs welfare", is stated as the following optimization:
\begin{eqnarray}
&& \min_{p,\rho} \mathbb{E}_\rho\left[\sum_{t=0}^{T-1}\sum_\alpha \left(\underbrace{U_\alpha(t+1)}_{\mbox{Cost of Electricity}}+\underbrace{\sum_\beta \log \frac{p_{\alpha\beta}(t)}{\bar{p}_{\alpha\beta}}}_{\mbox{welfare penalty}}\right)\right]_{\mbox{Eqs.~(\ref{stochastic},\ref{master-eq},\ref{rho_init})}},\label{profit_vs_welfare}\\
&=&{\min_{p,\rho} \sum_{t=0}^{T-1}\sum_\beta \rho_\beta(t)\left(\sum_\alpha p_{\alpha\beta}(t)\left(U_\alpha(t+1)+\log \frac{p_{\alpha\beta}(t)}{\bar{p}_{\alpha\beta}}\right)\right)_{\mbox{Eqs.~(\ref{stochastic},\ref{master-eq},\ref{rho_init})}}},
\end{eqnarray}
where the matrix $p$ is the optimization/control variable, which is stochastic according to Eq.~(\ref{stochastic}). Here in Eq.~(\ref{profit_vs_welfare}),
$\bar{p}$ is an exogenously known stochastic matrix describing the transition probabilities corresponding to normal dynamics/mixing within the ensemble, i.e., $\bar{p}$ explains the dynamics that the ensemble would show in the case of ``cost ignored" objective ($U=0$). The stochasticity of $\bar{p}$ means that $\bar{p}$ satisfies Eq.~(\ref{stochastic}), when $p$ is replaced by $\bar{p}$.

We assume that when following $\bar{p}$, {the ensemble mixes sufficiently fast to reach statistical steady state}, {$\rho^{(\mbox{\small st})}$}, i.e., $\bar{p} \rho^{(\mbox{\small st})}=\rho^{(\mbox{\small st})}$.

The essence of the optimization (\ref{profit_vs_welfare}) is a compromise that an aggregator aims to achieve between cost savings for the ensemble and keeping the level of discomfort (welfare penalty) to its minimum. The two conflicting objectives are represented by the two terms in Eq.~(\ref{profit_vs_welfare}).

Optimization (\ref{profit_vs_welfare}) is rather general, whereas for the example of a specific discrete-time--discrete-space TCL, one sets $U_\alpha$ to non-zero for only the states $\alpha$ that represent the ``switch-on" states of the TCL.

Eq.~(\ref{profit_vs_welfare}) belongs to the family of the so-called LS-MDPs introduced in \cite{07Tod}, discussed in \cite{12DjEmo,13DjEmo,15Meyn}, and briefly described as a special case in Appendix \ref{app:DP}. Solution of Eqs.~(\ref{profit_vs_welfare}) is fully described by Eqs.~(\ref{DP-solv}, \ref{lin-solv}, \ref{u_final}).

According to the general description part of Appendix \ref{app:DP}, the problem is solved in two DP steps:
\begin{itemize}
\item \underline{Backward in time step.} Compute $p$ recursively by advancing backward in time according to Eqs.~(\ref{DP-solv}, \ref{lin-solv}), where $\gamma(t)$ is substituted by $1$, and starting from the final condition Eq.~(\ref{u_final}).
\item \underline{Forward in time step.} Reconstruct $\rho$ by running Eq.~(\ref{master-eq}) forward in time with the initial condition Eq.~(\ref{rho_init}).
\end{itemize}

\section{Differentiating States through a Penalty/Encouragement}
\label{sec:penalty}

The KL welfare penalty term in Eq.~(\ref{profit_vs_welfare}) is restrictive in terms of how the transitions between different states, and also those observed at different moments of time, compare to each other. To encourage/discourage or generally differentiate the transitions, one may weight the terms in the KL sum differently, thus introducing the $\gamma_{\alpha\beta}(t)$ factors:
\begin{eqnarray}
\min_{p,\rho} \mathbb{E}_\rho\left[\sum_{t=0}^{T-1} \sum_\alpha \left(\underbrace{U_\alpha(t+1)}_{\mbox{Cost of Electricity}}+\underbrace{\sum_\beta { \gamma_{\alpha\beta}}(t)\log \frac{p_{\alpha\beta}(t)}{\bar{p}_{\alpha\beta}}}_{\mbox{welfare penalty}}\right)\right]_{\mbox{Eqs.~(\ref{stochastic},\ref{master-eq},\ref{rho_init})}}.
\label{penalty_opt}
\end{eqnarray}
This generalization of Eq.~(\ref{profit_vs_welfare}) aims to ease the implementation of the optimal decision, e.g., emphasizing or downplaying the controllability of transitions between the states and at different moments of time.

Solution of the optimization (\ref{penalty_opt}) is described in Appendix \ref{app:DP}. Even though linear solvability of the state-uniform formulation, discussed in Section \ref{sec:profit_optimal}, does not extend to the non-uniform formulation of Eq.~(\ref{penalty_opt}), the basic DP approach still holds and the problem is solved via the following backward-forward algorithm:
\begin{itemize}
\item \underline{Backward in time step.} Starting with the final conditions Eq.~(\ref{varphi_final}), one solves \\ Eq.~(\ref{varphi-def}) for $\varphi$ recursively backward in time. Solving Eq.~(\ref{varphi-def}) requires, at each time step, execution of inner loop iterations (until convergence), according to Eq.~(\ref{lambda_iterative}), to find the Lagrangian multiplier $\lambda$. {Following it, $p$ at that time is reconstructed according to Eq.~(\ref{DP-gen-1}). Alternatively, $p$ can be determined by minimizing convex function Eq.~(\ref{varphi-def}) directly with a linear constraint reflecting the stochasticity of $p$.}
\item \underline{Forward in time step.} Reconstruct $\rho$ running Eq.~(\ref{master-eq}) forward in time with the initial condition Eq.~(\ref{rho_init}).
\end{itemize}

\section{Computational Experiments}
\label{sec:experiments}

\begin{figure}
\centering
\includegraphics[scale=0.5]{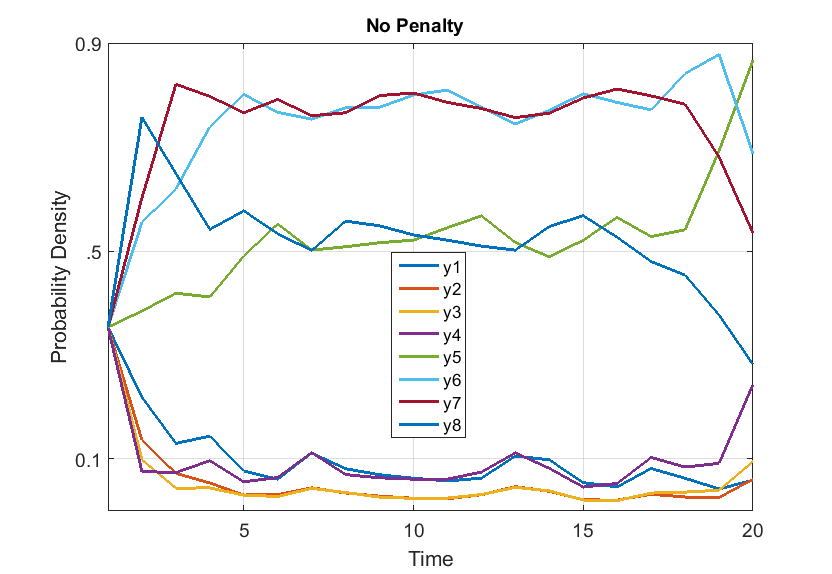}
\caption{Example solution of the MDP problem for the case without penalty, i.e., with the objective represented by Eq.~(\ref{profit_vs_welfare}). The initial probability distribution corresponds to the steady state of the ``target" MC with the transition probabilities shown in Fig.~(\ref{fig:TCL_MDP}). Eight curves show optimal dynamics of the respective, $\rho_\alpha(t),\quad \alpha=1,\cdots,8$.
\label{fig:no_penalty}}
\end{figure}

\begin{figure}
\centering
\includegraphics[scale=0.5]{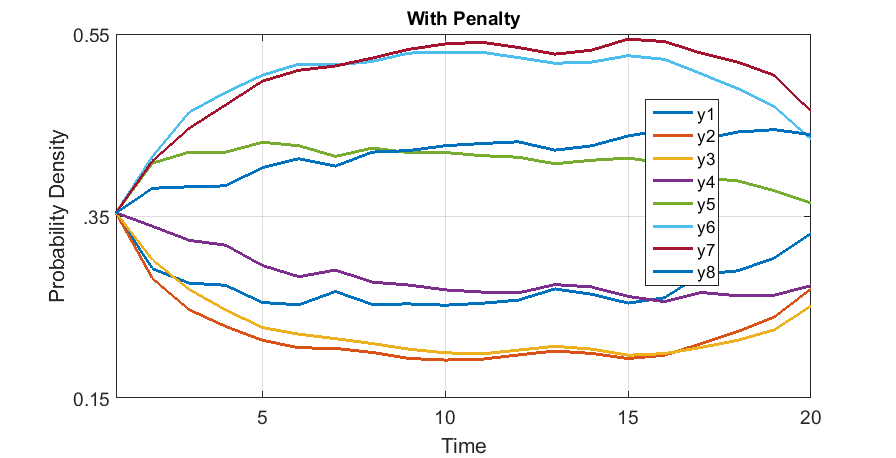}
\caption{Example solution of the MDP problem for the case with penalty, i.e., with the objective represented by Eq.~(\ref{penalty_opt}). The initial probability distribution corresponds to the steady state of the ``target" MC with the transition probabilities shown in Fig.~(\ref{fig:TCL_MDP}). Eight curves show optimal dynamics of the respective, $\rho_\alpha(t),\quad \alpha=1,\cdots,8$ for the setting, equivalent to the one used in Fig.~(\ref{fig:no_penalty}).
\label{fig:with_penalty}}
\end{figure}

In this section we describe some (preliminary) computational experiments conducted for MDP settings discussed in the two preceding sections. We choose the example case with $8$ states (four ``on", four ``off") and target transition probability, $\bar{p}$, in the form shown in Fig.~(\ref{fig:TCL_MDP}). We conduct the experiments in the regime with a non-stationary, time-dependent, and random cost term, $\sim 1+\mbox{rand}(t)$, that is nonzero for only the first four (``on") states. For the same target $\bar{p}$, we show solutions of two optimization formulations that correspond to the setting of Eq.~(\ref{profit_vs_welfare}) and Eq.~(\ref{penalty_opt}), respectively. In the latter case, we choose a time-independent penalty factor $\gamma_{\alpha\beta}=10$ for all transitions except for those that correspond to advancing one (counterclockwise) step along the cycle. The special ``along the cycle" transitions are not penalized and given a penalty factor equal to unity, $\gamma_{\mbox{mod}[\alpha+1],\alpha}=1,\quad \alpha=1,\cdots,7$. We use the algorithms described at the end of Section \ref{sec:profit_optimal} and Section \ref{sec:penalty}, respectively to solve the MDPs.

The results are shown in Figs.~(\ref{fig:no_penalty}, \ref{fig:with_penalty}). Comparing the figures corresponding to the two regimes (with and without penalty), one observes that imposing penalty leads to a more homogeneous distribution $\rho_\alpha$ over the states $\alpha$.

\section{Ancillary Services-vs-Welfare Optimal}
\label{sec:freq_control}

Another viable business model for an aggregator of an ensemble of cycling devices is to provide ancillary (frequency control) services to a regional SO. The ancillary services consist of adjusting the energy consumption of the ensemble to an exogenous signal. Tracking the signal may require some (or all) participants of the ensemble to sacrifice their natural cycling behavior. In this section we aim to study whether a perfect tracking of a predefined (exogenous) signal is feasible and then, in the case of feasibility, we would like to find the optimal solution causing least discomfort to the ensemble. Putting it formally, the aggregator solves the following optimization problem
\begin{eqnarray}
&\min_{p}& \mathbb{E}_\rho\left[\sum_{t=0}^{T-1} \underbrace{\sum_{\alpha,\beta} \gamma_{\alpha\beta}(t)\log \frac{p_{\alpha\beta}(t)}{\bar{p}_{\alpha\beta}}}_{\mbox{welfare penalty}}\right]\label{frequency_control}\\
&\mbox{s.t.}& \qquad \mbox{Eqs.~(\ref{stochastic})}\nonumber\\
&& \underbrace{s({t})=\sum_\alpha \varepsilon_\alpha \rho_\alpha({t}),\quad \forall \alpha,\quad \forall {t}=1,\cdots,T}_{\mbox{energy tracking constraint}}
\label{energy_tracking}
\end{eqnarray}
where $\bar{p}$ is the ``target" distribution represented by a stochastic matrix, e.g., one that leads to the standard steady state when $U=0$. $p$ is the stochastic matrix which is constrained by Eq.~(\ref{stochastic}). $s(t)$ is the amount of energy requested by the system operator to balance the (transmission level) grid, and $\varepsilon_\alpha$ is the amount of energy consumed by a device when it stays in the state $\alpha$ for a unit time slot. The setting of Eq.~(\ref{energy_tracking}) assumes perfect tracking, that is, the total consumption of the ensemble is exactly equal to the amount requested by the SO.

Introducing the Lagrangian multiplier for the energy tracking constraint, one restates Eq.~(\ref{frequency_control}) as the following min-max {(max-min according to normal Lagrangian formulation)} optimization:
\begin{eqnarray}
{\max\limits_\xi\min\limits_{p,\rho}} \!\left(\!\!\mathbb{E}_\rho\left[\sum\limits_{t=0}^{T-1} \sum\limits_{\alpha,\beta} \gamma_{\alpha\beta}(t)\log \frac{p_{\alpha\beta}(t)}{\bar{p}_{\alpha\beta}}
\!+\!\sum\limits_{t=1}^{T}\sum\limits_\alpha \xi(t)\varepsilon_\alpha\right]\!-\!\sum_{t=1}^{T} \xi(t)s(t)\!\right)_{
\mbox{Eq.~(\ref{stochastic})}},
\label{frequency_control_Lagr}
\end{eqnarray}
We observe that optimization over $p$ and $\rho$ in the resulting expression becomes equivalent under a substitution
\begin{eqnarray}
U_\alpha({t})=\xi({t})\varepsilon_\alpha,\quad \forall \alpha,\quad {t}=1,\cdots,T.
\label{U_via_xi}
\end{eqnarray}
to the penalized KL welfare penalty optimization (\ref{penalty_opt}), discussed in Section \ref{sec:penalty}.

In spite of the close relation between the energy tracking problem Eq.~(\ref{energy_tracking}) and the profit optimality problem Eq.~(\ref{penalty_opt}), the former is more difficult to implement. Indeed one can solve Eq.~(\ref{penalty_opt}) in only one backward-forward run. On the other hand, we are not aware of the existence of a similar efficient algorithm for solving Eq.~(\ref{energy_tracking}). The difficulty is related to the fact that $\xi(t)$ itself should be derived as the result of a KKT condition that reinforces the energy tracking constraint Eq.~(\ref{energy_tracking}). A natural resolution of this problem is through an outer-loop iteration including the following two substeps:
\begin{itemize}
\item Run the backward-forward penalty optimization algorithm described at the end of Section \ref{sec:penalty} using the current $\xi(t)$ (outer-step-specific) profile.
\item Update current $\xi(t)$ according to Eq.~(\ref{energy_tracking}), utilizing the current $\rho(t)$ derived from the previous substep.
\end{itemize}
The outer-loop iterative scheme is initiated with $\xi(t)$ derived from Eq.~(\ref{energy_tracking}), with $\rho(t)$ corresponding to the ``normal" MP, $p\to\bar{p}$. Iterations are run until a preset tolerance is achieved. We plan to experiment with and analyze convergence of this iterative scheme in the future.

\section{Hybrid Modeling: Toward Voltage-Aware \& Hierarchical Ensemble Control}
\label{sec:voltage}

\begin{figure}
\centering
\includegraphics[scale=0.35,page=3]{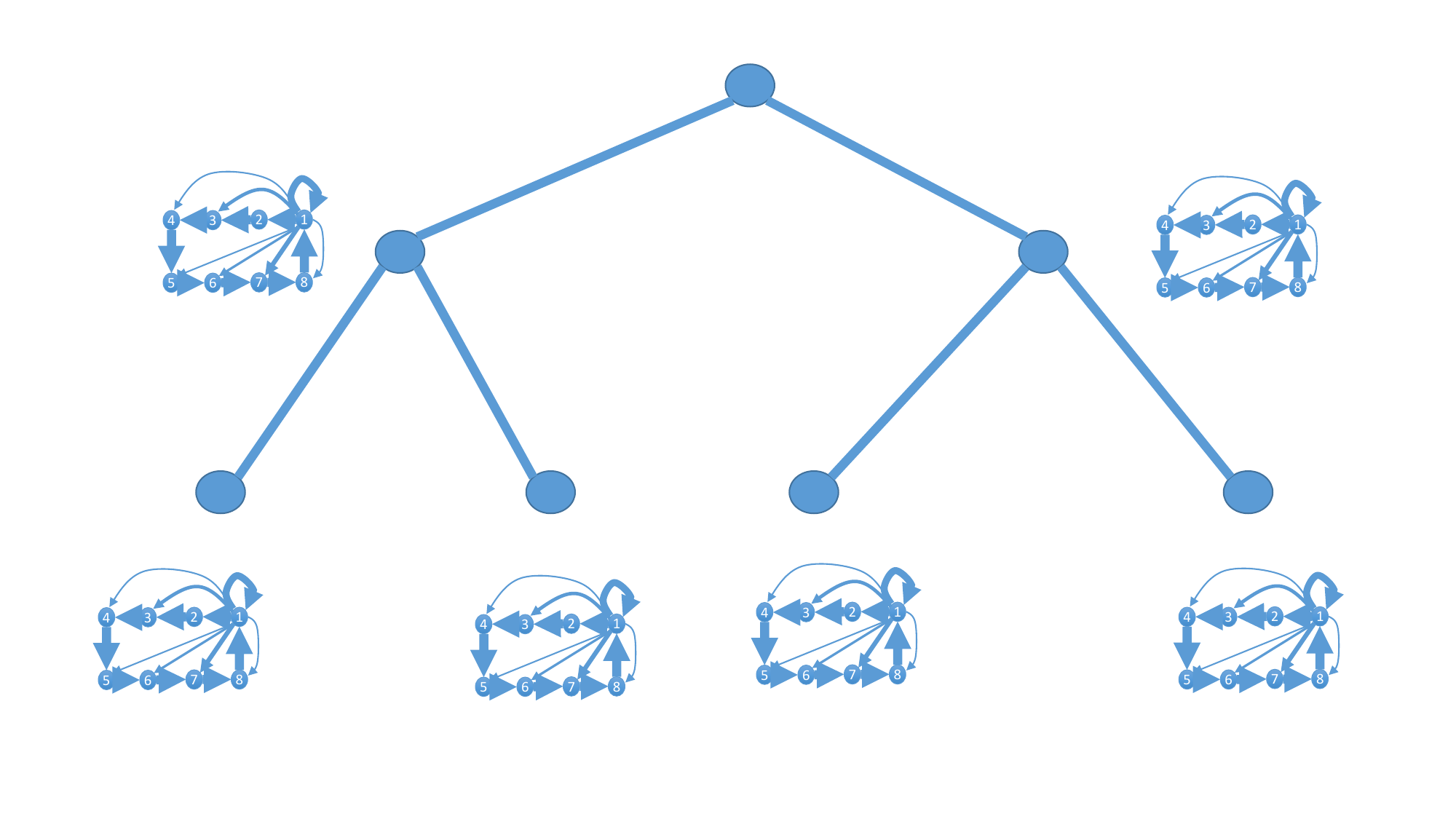}
\caption{Scheme illustrating hybrid/integrated and power-grid-aware MDP formulation. Nodes of the distribution level power system are each characterized by a  sub-ensemble modeled as an MP. Nodes are connected in a tree-like power system.}
\label{fig:MDP+grid}
\end{figure}

In this section we describe one possible scheme of an MDP approach for integration into control of power distribution networks. The description here is preliminary and meant to motivate a formulation for further exploration.

The aggregation of many loads discussed in the manuscript so far has ignored details of power flows (PF) as well as related voltage and line flow  constraints. We have assumed that any of the discussed solutions is PF-feasible, i.e. solution of PF equations is realizable for any of the consumption configurations with voltages and line flows staying within the prescribed bounds. For these assumptions to hold, either the aggregated devices should be in immediate geographical proximity of one another or the power system should be operated with a significant safety margin. In general, satisfying either of the two conditions globally for a large ensemble is impractical. This motivates a discussion of the following hybrid model, stated in terms of many geographically localized and different sub-ensembles connected into a power distribution network (see also Fig.~(\ref{fig:MDP+grid}) for illustration):
\begin{eqnarray}
&\min\limits_{p,\rho,s,V}& {\mathbb{E}_\rho}\left[\sum\limits_{t=0}^{T-1}
{\sum\limits_i}\sum\limits_\alpha \left(U_\alpha^{(i)}(t+1)+\sum\limits_\beta \gamma_{\alpha\beta}^{(i)}(t)\log \frac{p_{\alpha\beta}^{(i)}(t)}{\bar{p}_{\alpha\beta}^{(i)}}\right)\right]
\label{hybrid}\\
&s.t.&\nonumber\\
&& \rho_\alpha^{(i)}(t+1)=\sum_\beta p_{\alpha\beta}^{(i)}(t)\rho_{\beta}^{(i)}(t),\quad \forall t,\quad \forall i,\quad \forall \alpha
\label{hybrid-master}\\
&& \underbrace{s^{(i)}(t)\doteq \sum_\alpha \varepsilon^{(i)}_\alpha \rho_\alpha^{(i)}{(t)} =V^{(i)}(t)\sum_{j\sim i}\left(\frac{V^{(i)}(t)-V^{(j)}(t)}{z_{ij}}\right)^*}_{\mbox{Nodal PF relations}},\quad \forall i,~\forall t,
\label{hybrid-PF}\\
&& \underbrace{\underline{v}^{(i)}\leq |V^{(i)}|(t)\leq \overline{v}^{(i)}}_{\mbox{voltage constraints}},\quad \forall i,\quad \forall t.
\label{hybrid-voltage}
\end{eqnarray}
where each sub-ensemble $i$, representing for example stochastic/fluctuating consumption of a large apartment complex, is modeled as an aggregated MP that includs devices and consumers of different types. The objective (\ref{hybrid}) generalizes Eq.~(\ref{penalty_opt}), thus accounting for the cost of services and welfare quality (deviation from normal) for multiple ensembles. As done earlier in the manuscript, we introduce the additional penalty factor, $\gamma^{(i)}_{\alpha{\beta}}(t)$, in front of the KL (welfare penalty) term. This allows us to reflect the significance of different ensembles, transitions, and times.
Eq.(\ref{hybrid-master}) generalizes the ME (\ref{master-eq}) to sub-ensembles.
Eqs.~(\ref{hybrid-PF}, \ref{hybrid-voltage}) introduce PF and voltage constraints into the MDP setting. We thus seek for an optimum (\ref{hybrid}) over
three vectors: the vector $ p=\left(p_{\alpha\beta}^{(i)}(t)\left|\forall i,\ \ \forall \alpha,\beta,\ \ \forall t\right.\right)$, constructed from the stochastic transition-probability matrices, where each component represents a node in the network at all moments of time; the vector $\rho=\left(\rho_{\alpha}^{(i)}(t)\left|\forall i,\ \ \forall \alpha,\ \ \forall t\right.\right)$, reconstructed from $p$ according to Eqs.~(\ref{hybrid-master}) with the proper initial conditions provided; and the vector of voltages $ V=\left(V^{(i)}(t)\left|\forall i,\ \ \forall t\right.\right)$.

Notice that other objectives (such as a contribution enforcing minimization of power losses in the distribution systems), other constraints (such as imposing bounds on line flows), as well as other controls (such as voltage/ position of tap-changers and/or reactive consumption at the nodes containing inverters) can be incorporated into the scheme in the spirit of \cite{11TSBC,14SBC}.

Because the MDP problem is stochastic by nature, it also allows incorporation of other stochastic sources, e.g., solar or wind renewables, which can be done either via modeling the stochastic sources as MP (with no control) or by extending the model by adding so-called chance-constrained descriptions in the spirit of \cite{14BCH}.

An efficient solution of the hybrid problem (\ref{hybrid}) can be built by combining the techniques of temporal DP, developed in this manuscript for individual MDPs, with spatial (tree-graph) DP techniques, developed recently for the Optimal PF (OPF) in power distributions \cite{Dvijotham2016} and taking advantage of the tree-graph operational layout of power distribution networks. We plan to work on practical implementation of these and other components of the hybrid model in the future.

\section{Conclusions \& Path Forward}
\label{sec:conclusions}

In this manuscript we review and develop a computationally scalable approach for optimization of an ensemble of devices modeled via finite-space, finite-time MP. This approach builds upon earlier publications \cite{15GMK,15MBBCE,15PKL} addressing DR applications in power systems and beyond. A particularly practical and popular example of MP, relevant for power systems, is the ensemble of cycling loads, such as air conditioners, water heaters, or residential water pumps \cite{11CH,14Siano}.

We have developed a number of useful MDP formulations aiming to achieve optimality for a diverse set of objectives of interest for an aggregator (of the ensemble) under different circumstances. We started the manuscript by describing MDP that balances overall expenses of the ensemble acquired when the cost of electricity varies in time, with a welfare penalty that measures ensemble operational deviation from its normal behavior. Then we proceeded to discuss MDP built to test feasibility of the ensemble to provide ancillary frequency control services to SO. Finally, we addressed the future challenge of building a hybrid model that incorporates MP and MDP modeling into deterministic and stochastic frameworks of OPFs (operational dispatch) for power distribution.

The most important technical achievement of this paper is the development of an appropriate DP framework for stochastic optimization including DR. For the cost-vs-welfare optimization we also tested the developed methodology and algorithms numerically (on a small-scale example).

This manuscript does not offer a concluding fine-tuned summary of a completed project. Instead, we have focused here on presenting a new open-ended paradigm. In other words, we expect to see emergence of many more future extensions and generalizations of the approach that we have started to develop here. Some of these proposed future developments have been already discussed in the preceding sections, especially in Section \ref{sec:voltage}. Others are briefly highlighted below.
\begin{itemize}
\item {\it Utilizing and Extending Lin-Solvability.} Intuitively, it is clear that lin-solvability is a rich property that should be advantageous for both analysis and algorithms. However, the use of this strong property in this manuscript was rather limited. We expect that the lin-solvability will provide an actual computational/algorithmic benefit not just for solving MDP per se but also for solving more-complex multilevel optimization/control problems where an individual MDP represents an element of a richer model. (Some examples are mentioned below.) On the other hand, the lin-solvability described in Appendix \ref{app:subsec_linear} is a rather delicate property of the model and it is easy to lose either partially or completely, as shown for the setting/formulation discussed in Appendix \ref{app:subsec_norm} and the main part of Appendix \ref{app:DP}, respectively. This observation motivates further investigation of other settings/formulations amenable for either full or partial lin-solvability. One promising direction for future analysis is to consider a generalization from the KL-divergence to Renyi-divergence, building upon and extending the approach of \cite{12DjEmo,13DjEmo}.

\item {\it Model Reduction.} Many models of power systems are too large and detailed for computations. This applies even to the routine PF computation, which is a routing subtask for many power system problems of high level involving optimization, control, and generalizations accounting for stochasticity and robustness. Seeking to represent large-scale, long-time behavior, one is interested in building a reduced model in order to aggregate the small-scale and/or short-time details in a compact way. The reduction may be lossless or lossy, with or without the ability to reconstruct the small-scale/short-time details. The intrinsically stochastic MP framework developed here is appropriate for the lossy case, where stochasticity represents the loss of insignificant details. We envision building reduced models capable of more efficient but still accurate computations of, e.g., PF, in the format similar to that discussed in Section \ref{sec:voltage}. The reduced model may consist of power lines with effective impedances and stochastic load/generation elements represented by MPs. The level of coarse-graining may be predefined by a geography-preserving procedure, e.g., of the type discussed in \cite{17GDCB}, but it may also be left flexible/adaptive, where the number of states and allowable transitions for an individual MP is the subject of optimization.

\item {\it Hierarchical Control.} When the reduced model with MPs representing aggregated loads/generators is to be used in the context of optimization and control, changing from MP to MDP may be understood as allowing the optimization/control to be split into two levels. Optimization with many MDPs modeling aggregated end-users, as discussed in Section \ref{sec:voltage}, will produce optimal transition probability matrices for each individual MDP. Then, the task of implementing this policy is delegated to aggregators responsible for individual MDPs. In this case, implementation means the global optimality/control is substituted by a two-level hierarchical control. The scheme may be extended to represent more levels of control.

\item {\it Derivation of MDP from a Detailed Microscopic Model.} Consider an ensemble of continuous-time TCLs, each represented by a microscopic state in terms of temperature (continuous variable) and switch on/off status (discrete variable). In the case of an inhomogeneous ensemble, where each TCL or a group of TCLs may be parameterized differently, e.g., in terms of the allowed temperature rates and/or transition rates between on/off states, one is posing the question of representing this inhomogeneous ensemble as a single discrete time and discrete space (binned) MP or MDP. Developing a methodology for constructing a representative MP/MDP for an aggregated ensemble is an important task for future research.

\item {\it Accounting for Risk-Metric.} The MP/MDP methodology is sufficiently rich and flexible to account for and mitigate risks of different types, e.g., overloading. Indeed, incorporating additional constraints into MDP optimization by limiting some elements of $\rho$ or $p$ is a straightforward way of limiting the risk in probabilistic terms, natural for MDP.

\item {\it Stochastic MDP.} MDP models already account for the intrinsic stochasticity of an ensemble of devices (or coarse-grained areas) that a model represents. However, exogenous effects, such as those representing the cost of electricity or the frequency signal, are modeled in this manuscript deterministically, even though it is often more appropriate to represent generically uncertain and stochastic exogenous signals probabilistically. Formulating appropriate ``second-order" statistical models represents an interesting challenge for future research. Interesting recent studies dealing with exogenous fluctuations and uncertainty added to the MP/MDP setting are presented in \cite{16BMa,16BMb}.

\item {\it From MDP to Reinforcement Learning (RL).} MDP is a standard tool used in the field of RL. One of the data-driven RL approaches \cite{07Tod} relevant for our discussion suggests considering ``ideal" transition probabilities, $\bar{p}$, as unknown/uncertain and then attempting to learn $\bar{p}$ from the data in parallel with solving the MDP. The approach was coined in \cite{07Tod} as $Z$-learning, and it is also closely related to the so-called approximate DP (see \cite{RL} and references therein for many books and reviews). We anticipate that this general approach, when applied to the models introduced in the manuscript, will allow us to build a data-driven and MDP-based framework for controlling an ensemble whose normal behavior is known only through limited samples of representative behavior.

\item {\it MDP for Supervised Learning (SL).} Recently a SL methodology was applied as a real-time proxy to solve difficult power system problems, such as finding an efficient description of the feasibility domain \cite{17DHMT} to solve OPF \cite{16CDM} or power system reliability management problems \cite{16Duc}. The main idea in this line of research rests on replacing an expensive power system computation with an ML black box trained to evaluate a sufficient number of samples labeled by relevant output characteristics. One interesting option consists of using properly designed MPs and/or MDPs for opening up the black box and turning it into (at least partially) a physics-informed ML. Another possible direction would be to use MPs/MDPs as labels.

\item{\it MDP as an Element of a Graphical Model (GM) framework.} It was argued recently in \cite{Dvijotham2016,17CKMVV} that the approach of GM offers a flexible framework and efficient solutions/algorithms for a variety of optimization and control problems in energy systems (power systems and beyond). It is of interest to extend the GM framework and build into it MDP methodology, formulations, and solutions.

\item {\it Applications to Systems, e.g., Energy Systems.} The MDP models discussed in this manuscript are of interest well beyond representing ensembles (and coarse regions) in power systems. Similar methods and approaches are relevant for problems representing behavior and DR capabilities of consumers'/producers' ensembles in other energy infrastructures, such as natural gas systems and district heating systems. In fact, the models discussed above fit practically ``as is" to describe aggregation of many consumers of district heating systems residing in a big apartment complex or a densely populated residential area. The language of MDP is also universal enough to optimize joint energy consumption through multiple energy infrastructures of a ``residential" ensemble (including electricity, heat, and gas---possibly used as an alternative to central heating in local boilers).
\end{itemize}

\section{Acknowledgments}

The authors are grateful to S. Backhaus, I. Hiskens, and D. Calloway for fruitful discussions, guidance, and valuable comments. The work at LANL was carried out under the auspices of the National Nuclear Security Administration of the U.S. Department of Energy under Contract No. DE-AC52-06NA25396.

\appendix

\section{Dynamic/Bellman Programming}
\label{app:DP}

In this appendix we discuss the DP solution for the most general of the formulations considered in this paper, the profit-vs-welfare optimal formulations. Specifically we consider Eq.~(\ref{penalty_opt}) that introduces in-homogeneity in the inter-state transitions, expressed through the state and time-dependent $\gamma_{{\beta}\alpha}(\tau)$ factors.

Let us restate Eq.~(\ref{penalty_opt}) as
\begin{eqnarray}
\min_{p}\ C\left(\left.\left. p\right|_0^{T-1},\rho(0)\right)\right|_{\sum_\beta p_{{\beta}\alpha}(\tau)=1,\ \forall \alpha},
\label{penalty_opt1}
\end{eqnarray}
where $\left. p\right|_0^{T-1}$ is the shortcut notation for the vector constructed of the transition probability matrices evaluated at $t=0,\cdots,T-1$,
i.e.,
 $(p(t)|\forall t=0,\cdots, T-1{)}$. The so-called value function in Eq.~(\ref{penalty_opt1}) can be decomposed according to
\begin{eqnarray}
\hspace{-1cm} C\left(\left. p\right|_0^{T-1},\rho(0)\right)=C\left(\left. p\right|_0^{\tau-1},\rho(0)\right)+
\sum_\alpha \varphi_\alpha\left(\tau, \left. p\right|_\tau^{T-1}\right)\rho_\alpha\left(\tau,\left. p\right|_0^{\tau-1}\right),
\label{C_decomposed}
\end{eqnarray}
where the notation $\rho_\alpha\left(\tau,\left. p\right|_0^{\tau-1}\right)$ is introduced temporarily (just for the purpose of this derivation) to emphasize that $\rho_\alpha$ computed at the moment of time $\tau$ also depends, according to
Eq.~(\ref{master-eq}), on the transition probability matrices computed at all the preceding times. Here in Eq.~(\ref{C_decomposed}) $\varphi_\alpha$ is defined at the final moment of time according to
\begin{eqnarray}
\varphi_\alpha(T)=U_\alpha(T),
\label{varphi_final}
\end{eqnarray}
and then solved backward in time by the following recursive equations
\begin{eqnarray}
&& \forall \alpha,\quad \tau=T-1,\cdots 0:\quad \varphi_\alpha\left(\tau,\left. p\right|_\tau^{T-1}\right)=\sum_\beta \varphi_\beta\left(\tau+1,\left. p\right|_{\tau+1}^{T-1}\right)p_{\beta\alpha}(\tau)\nonumber\\
&& + \sum_\beta \gamma_{\beta\alpha}(\tau) p_{\beta\alpha}(\tau)\log\left(\frac{p_{\beta\alpha}(\tau)}{\bar{p}_{\beta\alpha}}\right)+U_\alpha(\tau),
\label{varphi-def}
\end{eqnarray}
where, the notation $\varphi_\alpha\left(\tau,\left. p\right|_\tau^{T-1}\right)$ emphasizes that (by construction) $\varphi_\alpha(\tau)$ depends only on $\left.p\right|_\tau^{T-1}$.

The DP-decomposed (recursive) structure of Eqs.~(\ref{C_decomposed}, \ref{varphi-def}) allows us to evaluate optimization over $p$ in Eq.~(\ref{penalty_opt1}) greedily as Karush–Kuhn–Tucker (KKT) first-order conditions---first over $p(T-1)$ and then over $p(T-2)$ all the way to $p(0)$. {The KKT condition for minimization of (\ref{varphi-def}) with linear constraint $\sum_\beta p_{\beta\alpha}(\tau)=1$ gives the following DP relation for the optimal $p$:
\begin{eqnarray}
\hspace{-1cm} p_{\beta\alpha}(\tau)=\bar{p}_{\beta\alpha}\exp\left(-1-\frac{\varphi_\beta(\tau+1)-\lambda_\alpha(\tau)}{\gamma_{\beta\alpha}(\tau)}\right),\ \forall\alpha,\beta,\ \forall \tau=T-1,\cdots,0,
\label{DP-gen-1}
\end{eqnarray}
where the Lagrange multipliers, $\lambda_\alpha(\tau)$, are determined implicitly from the stochasticity of $p(\tau)$
\begin{eqnarray}
\sum_\beta \bar{p}_{\beta\alpha}\exp\left(-1-\frac{\varphi_\beta(\tau+1)-\lambda_\alpha(\tau)}{\gamma_{\beta\alpha}(\tau)}\right)=1,\ \forall \alpha,\ \forall \tau=T-1,\cdots,0.
\label{lambda_implicit}
\end{eqnarray}
The Lagrange multipliers $\lambda$ can also be extracted from one-dimensional convex optimizations one gets substituting  (\ref{DP-gen-1}) into the corresponding Lagrangian relaxation of (\ref{varphi-def})
\begin{eqnarray}
&& \lambda_\alpha(\tau)=\mbox{arg}\min_\mu\left(
\sum_\beta \bar{p}_{\beta\alpha}\gamma_{\beta\alpha}(\tau)\exp\left(-1-\frac{\varphi_\beta(\tau+1)-\mu}{\gamma_{\beta\alpha}(\tau)}\right)-\mu\right),
\label{lambda_opt}\\
&& \forall \alpha,\ \forall \tau=T-1,\cdots,0.
\nonumber
\end{eqnarray}
Computationally, one may solve Eq.~(\ref{lambda_opt}) via gradient descent
\begin{eqnarray}
&& \lambda_\alpha(\tau)\leftarrow \lambda_\alpha(\tau)-
\delta \Biggl(\sum_\beta \bar{p}_{\beta\alpha} \exp\left(-1-\frac{\varphi_\beta(\tau+1)-\lambda_\alpha(\tau)}{\gamma_{\beta\alpha}(\tau)}\right)-1\Biggr),
\label{lambda_iterative}\\ && \forall \alpha,\ \forall \tau=T-1,\cdots,0.
\nonumber
\end{eqnarray}
choosing an appropriate step, $\delta$, and iterating Eq.~(\ref{lambda_iterative}) until the target tolerance of the solution's accuracy is reached. Then Eqs. (\ref{DP-gen-1}, \ref{varphi-def}) can be used to determine $p(\tau)$ and $\varphi(\tau)$.\\ Another way to determine $p(\tau),\varphi(\tau)$ is to minimize (\ref{varphi-def}) directly resolving the $p$-stochasticity constraint. As $x\log x$ is a convex function, minimizing (\ref{varphi-def}) is a convex problem that can be solved with standard solvers, e.g. Ipopt or cvx.}

\subsection{The case which allows explicit normalization}
\label{app:subsec_norm}

In the case of a general $\gamma(\tau)$ Eq.~(\ref{DP-gen-1}) the Lagrange multipliers, $\lambda(t)$, cannot be expressed via $p$ in a closed form. An exception is the case when $\gamma_{\beta\alpha}(\tau)$ is independent of $\beta$, i.e.
{\begin{eqnarray}
\gamma_{\beta\alpha}(\tau)\Rightarrow \gamma_\alpha(\tau).
\label{closed-form}
\end{eqnarray}}
Then, Eq.~(\ref{DP-gen-1}) results in
\begin{eqnarray}
p_{\beta\alpha}(\tau)=\frac{\exp\left(-\frac{\varphi_\beta(\tau+1)}{\gamma_\alpha(\tau)}\right)\bar{p}_{\beta\alpha}}{\sum_\nu \exp\left(-\frac{\varphi_\nu(\tau+1)}{\gamma_\alpha{(\tau)}}\right)\bar{p}_{\nu\alpha}},
\quad \forall \alpha,\beta,\quad \forall t.
\label{DP-solv}
\end{eqnarray}
Substituting Eq.~(\ref{DP-solv}) into Eq.~(\ref{varphi-def}), one derives
\begin{eqnarray}
&& \hspace{-1cm}\varphi_\alpha(\tau)=-\gamma_\alpha(\tau)\log\left(\sum_\beta
\exp\left(-\frac{\varphi_\beta(\tau+1)}{\gamma_\alpha(\tau)}\right)\bar{p}_{\beta\alpha}\right)+
U_\alpha(\tau),\quad \forall \alpha,\quad \forall \tau.
\label{varphi-DP-norm}
\end{eqnarray}

Notice that there are also other special cases that may allow for analytic expressions for the normalization. In particular, the case when $\gamma_{\alpha\beta}(\tau), \forall \tau,\alpha,\beta$ take values from a finite alphabet. We leave discussion of this and other interesting cases to future studies.

\subsubsection{Linearly solvable case}
\label{app:subsec_linear}

Reduction (\ref{closed-form}) allows us to map Eq.~(\ref{penalty_opt1}) to solution of DP equations (\ref{varphi-DP-norm}) that are, however, nonlinear. We now make an additional reduction to limit Eq.~(\ref{closed-form}) even further to the state independent case
\begin{eqnarray}
\gamma_{\alpha\beta}(\tau)\Rightarrow \gamma(\tau),
\label{lin-solv-reduction}
\end{eqnarray}
Eq.~(\ref{varphi-DP-norm}) now reduces to what was coined in \cite{07Tod,12DjEmo,13DjEmo} as the linearly solvable MDP
\begin{eqnarray}
\exp\left(-\frac{\varphi_\alpha(\tau)}{\gamma({\tau})}\right)\doteq u_\alpha(\tau)=\sum_\beta u_\beta(\tau+1) \bar{p}_{\beta\alpha}\exp\left(-\frac{U_\alpha(\tau)}{\gamma({\tau})}\right).
\label{lin-solv}
\end{eqnarray}
Eq.~(\ref{lin-solv}) is solved backward in time with the final condition
\begin{eqnarray}
u_\alpha(T)=\exp\left(-\frac{U_\alpha(T)}{\gamma(T)}\right),\quad \forall \alpha.
\label{u_final}
\end{eqnarray}

Two remarks are in order. First, we note that Eq.~(\ref{lin-solv-reduction}) is of practical interest when one aims to change the relative importance of the welfare reinforcement vs. price balance in the optimization. Second, other linearly solvable cases, in addition to those described by Eq.~(\ref{lin-solv-reduction}), may exist. We postpone a more general discussion of a broader class of the linearly solvable cases as well as their practical utility, to future publications.

\bibliographystyle{spmpsci}
\bibliography{TCL}

\end{document}